\documentclass[twoside,twocolumn,9pt]{article}
\usepackage{extsizes}
\usepackage[super,sort&compress,comma]{natbib} 
\usepackage[version=3]{mhchem}
\usepackage[left=1.5cm, right=1.5cm, top=1.785cm, bottom=2.0cm]{geometry}
\usepackage{balance}
\usepackage{times,mathptmx}
\usepackage{sectsty}
\usepackage{graphicx} 
\usepackage{lastpage}
\usepackage[format=plain,justification=justified,singlelinecheck=false,font={stretch=1.125,small,sf},labelfont=bf,labelsep=space]{caption}
\usepackage{float}
\usepackage{fancyhdr}
\usepackage{fnpos}
\usepackage[english]{babel}
\addto{\captionsenglish}{%
  
}
\usepackage{array}
\usepackage{droidsans}
\usepackage{charter}
\usepackage[T1]{fontenc}
\usepackage[usenames,dvipsnames]{xcolor}
\usepackage{setspace}
\usepackage[compact]{titlesec}
\usepackage{hyperref}
\usepackage{epstopdf}%This line makes .eps figures into .pdf - please comment out if not required.

\definecolor{cream}{RGB}{222,217,201}

\begin{document}

\pagestyle{fancy}
\thispagestyle{plain}
\fancypagestyle{plain}{

%%%HEADER%%%
\fancyhead[C]{\includegraphics[width=18.5cm]{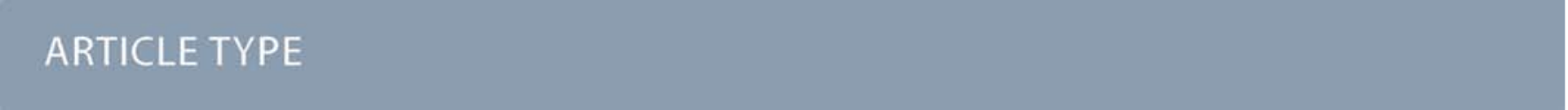}}
\fancyhead[L]{\hspace{0cm}\vspace{1.5cm}\includegraphics[height=30pt]{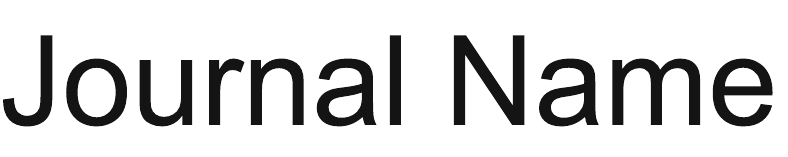}}
\fancyhead[R]{\hspace{0cm}\vspace{1.7cm}\includegraphics[height=55pt]{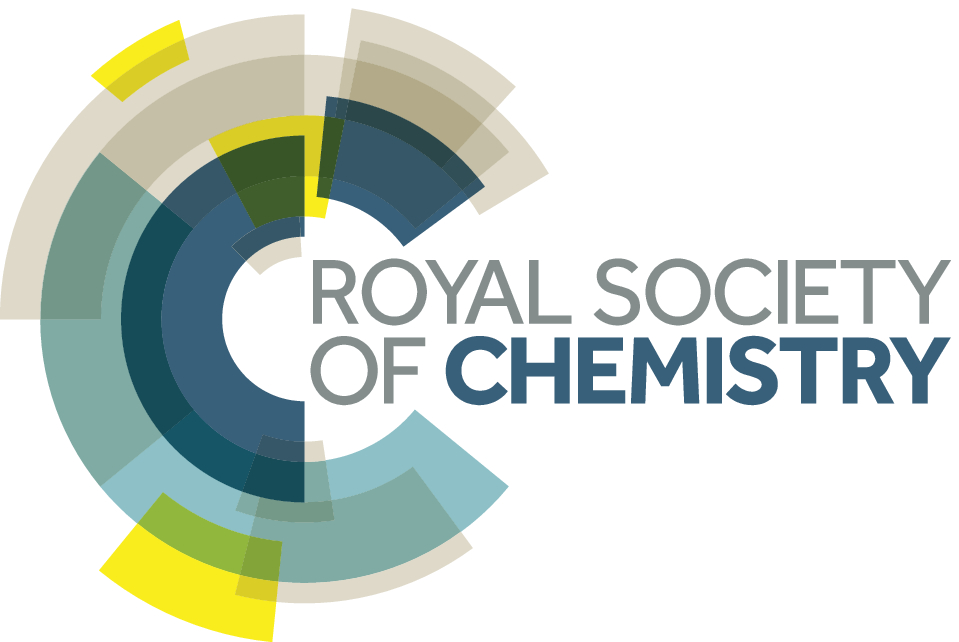}}
\renewcommand{\headrulewidth}{0pt}
}
%%%END OF HEADER%%%

%%%PAGE SETUP - Please do not change any commands within this section%%%
\makeFNbottom
\makeatletter
\renewcommand\LARGE{\@setfontsize\LARGE{15pt}{17}}
\renewcommand\Large{\@setfontsize\Large{12pt}{14}}
\renewcommand\large{\@setfontsize\large{10pt}{12}}
\renewcommand\footnotesize{\@setfontsize\footnotesize{7pt}{10}}
\makeatother

\renewcommand{\thefootnote}{\fnsymbol{footnote}}
\renewcommand\footnoterule{\vspace*{1pt}% 
\color{cream}\hrule width 3.5in height 0.4pt \color{black}\vspace*{5pt}} 
\setcounter{secnumdepth}{5}

\makeatletter 
\renewcommand\@biblabel[1]{#1}            
\renewcommand\@makefntext[1]% 
{\noindent\makebox[0pt][r]{\@thefnmark\,}#1}
\makeatother 
\renewcommand{\figurename}{\small{Fig.}~}
\sectionfont{\sffamily\Large}
\subsectionfont{\normalsize}
\subsubsectionfont{\bf}
\setstretch{1.125} %In particular, please do not alter this line.
\setlength{\skip\footins}{0.8cm}
\setlength{\footnotesep}{0.25cm}
\setlength{\jot}{10pt}
\titlespacing*{\section}{0pt}{4pt}{4pt}
\titlespacing*{\subsection}{0pt}{15pt}{1pt}
%%%END OF PAGE SETUP%%%

%%%FOOTER%%%
\fancyfoot{}
\fancyfoot[LO,RE]{\vspace{-7.1pt}\includegraphics[height=9pt]{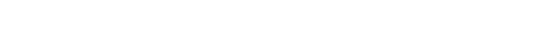}}
\fancyfoot[CO]{\vspace{-7.1pt}\hspace{13.2cm}\includegraphics{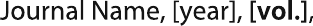}}
\fancyfoot[CE]{\vspace{-7.2pt}\hspace{-14.2cm}\includegraphics{head_foot/RF}}
\fancyfoot[RO]{\footnotesize{\sffamily{1--\pageref{LastPage} ~\textbar  \hspace{2pt}\thepage}}}
\fancyfoot[LE]{\footnotesize{\sffamily{\thepage~\textbar\hspace{3.45cm} 1--\pageref{LastPage}}}}
\fancyhead{}
\renewcommand{\headrulewidth}{0pt} 
\renewcommand{\footrulewidth}{0pt}
\setlength{\arrayrulewidth}{1pt}
\setlength{\columnsep}{6.5mm}
\setlength\bibsep{1pt}
%%%END OF FOOTER%%%

%%%FIGURE SETUP - please do not change any commands within this section%%%
\makeatletter 
\newlength{\figrulesep} 
\setlength{\figrulesep}{0.5\textfloatsep} 

\newcommand{\topfigrule}{\vspace*{-1pt}% 
\noindent{\color{cream}\rule[-\figrulesep]{\columnwidth}{1.5pt}} }

\newcommand{\botfigrule}{\vspace*{-2pt}% 
\noindent{\color{cream}\rule[\figrulesep]{\columnwidth}{1.5pt}} }

\newcommand{\dblfigrule}{\vspace*{-1pt}% 
\noindent{\color{cream}\rule[-\figrulesep]{\textwidth}{1.5pt}} }

\makeatother
%%%END OF FIGURE SETUP%%%

%%%TITLE, AUTHORS AND ABSTRACT%%%
\twocolumn[
  \begin{@twocolumnfalse}
\vspace{3cm}
\sffamily
\begin{tabular}{m{4.5cm} p{13.5cm} }

\includegraphics{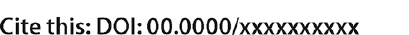} & \noindent\LARGE{\textbf{Detection of sub-degree fluctuations of the local cell membrane slope  using optical tweezers}} \\%Article title goes here instead of the text "This is the title"
\vspace{0.3cm} & \vspace{0.3cm} \\

 & \noindent\large{Rahul Vaippully,\textit{$^{a}$} Vaibavi Ramanujan,\textit{$^{b}$} Manoj Gopalakrishnan,\textit{$^{a}$} Saumendra  Bajpai,\textit{$^{b}$}and Basudev Roy\textit{$^{a}$}} \\%Author names go here instead of "Full name", etc.

\includegraphics{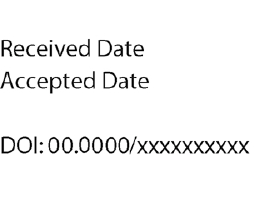} & \noindent\normalsize{Normal thermal fluctuations of the cell membrane have been studied extensively using high resolution microscopy and focused light, particularly at the peripheral regions of a cell. We use a single probe particle attached non-specifically to the cell-membrane to determine that the power spectral density is proportional to (frequency)$^{-\frac{5}{3}}$ in the range of 5 Hz to 1 kHz. We also use a new technique to simultaneously ascertain the slope fluctuations of the membrane by relying upon the determination of pitch motion of the birefringent probe particle trapped in linearly polarized optical tweezers. In the process, we also develop the technique to identify pitch rotation to a high resolution using optical tweezers. We find that the power spectrum of slope fluctuations is proportional to (frequency)$^{-1}$, which we also explain theoretically. We find that we can extract parameters like bending rigidity directly from the coefficient of the power spectrum particularly at high frequencies, instead of being convoluted with other parameters, thereby improving the accuracy of estimation. We anticipate this technique for determination of the pitch angle in spherical particles to high resolution as a starting point for many interesting studies using the optical tweezers. } \\%The abstract goes here instead of the text "The abstract should be..."

\end{tabular}

 \end{@twocolumnfalse} \vspace{0.6cm}

  ]
%%%END OF TITLE, AUTHORS AND ABSTRACT%%%

%%%FONT SETUP - please do not change any commands within this section
\renewcommand*\rmdefault{bch}\normalfont\upshape
\rmfamily
\section*{}
\vspace{-1cm}

%%%FOOTNOTES%%%

\footnotetext{\textit{$^{a}$ Department of Physics, Indian Institute of Technology Madras, Chennai, India, 600036 E-mail: basudev@iitm.ac.in}}
\footnotetext{\textit{$^{b}$ Department of Applied Mechanics, Indian Institute of Technology Madras, Chennai, India, 600036 }}

%Please use \dag to cite the ESI in the main text of the article.
%If you article does not have ESI please remove the the \dag symbol from the title and the footnotetext below.

%additional addresses can be cited as above using the lower-case letters, c, d, e... If all authors are from the same address, no letter is required

%%%END OF FOOTNOTES%%%

Rheology of the cell membrane assumes significance in cell migration, adhesion, differentiation and development \cite{parsons, lecuit,mcmahon,kim}, not to mention, also in probing the health of the cell. It is directly influenced in diseases like malaria \cite{park} and sickle cell anaemia \cite{connes}. Further, the cancer cells are softer and more elastic compared to healthy ones to help in intravasation \cite{wirtz}, when trying to get into the blood vessels and spread through the body, where the exact mechanism by which it changes the elasticity is not known \cite{katz}. In view of all these facets, study of membrane stiffness and the subsequent response to external perturbations assume enormous importance. 

Membrane fluctuations are inherent to many membrane processes, like ion-pump functioning, vesicle budding and trafficking \cite{gov,gov1,park1} in living cells. Our knowledge of the mechanisms of the membrane processes shall be significantly improved while learning about the nature of active fluctuations \cite{yu,biswas} in membranes. 

Typically, the normal membrane fluctuations have been studied to ascertain the rheological parameters of the living cells \cite{biswas}. These fluctuations are powered by thermal energy as well as ATP dependent processes. The temporal range of such fluctuations is quite broad, starting from slow (10 sec) actin waves to drive large wavelength fluctuations (100 nm to 10 $\mu$m) at cell edges and basal membrane \cite{giannone,dobereiner,chen}, to relatively smaller amplitude ones (5 to 50 nm) which appear at the basal membrane \cite{monzel,santos} and are mainly thermal in nature. Fluctuations of the basal membrane, as opposed to the cell edges have not been explored much due to requirements of high resolution. We use a new technique where we place a particle on top of a cell membrane at locations away from the cell edges to find the normal fluctuations after ensuring non-specific binding. This does not require proximity to a second surface as the interference is between the unscattered light in photonic force microscopy with that of the scattered light from the particle \cite{florin,rohrbach,rohrbach1}, and thus the unconfined free surface of the cell can also be probed. 

Here we introduce a hitherto new concept, that of membrane local slope fluctuations, to study the parameters. To perform such a measurement, we show how the pitch-rotation angle \cite{basudev} of a spherical particle attached to the membrane can be ascertained at high resolution in optical tweezers to add additional parameters that can greatly improve the accuracy. We show that such measurement provides information of the parameters like bending rigidity directly instead of being ascertained in a convoluted form with other parameters. In the process, we show for the first time, a method to determine the pitch rotation angle to a high resolution using optical tweezers. 

\section{Theory}
The pitch signal is given as the difference-in-halves signal of the light scattered by the birefringent particle placed inside crossed polarizers \cite{basudev}, and can also extend to  particles trapped in optical tweezers. The pitch signal is linearly proportional to the difference-in-halves signal. The power spectrum due to pitch Brownian motion is given as follows, in consistency with the conventional power spectra in optical tweezers \cite{erik}.

\begin{equation}
    PSD=\frac{A}{\omega^2+B}
    \label{lorentz}
\end{equation}\\
Further, following the Wiener-Khinchin theorem, the power spectral density (PSD) of membrane height fluctuations is given by

\begin{equation}
PSD_z=\int dt e^{i\omega t}\int \frac{d^2{\bf q}d^2{\bf q}^{\prime}}{(2\pi)^4}\langle h_{\bf q}(0)h_{{\bf q}^{\prime}}(t)\rangle
\label{wiener}
\end{equation}
where $h_{\bf q}(t)=\int d^2{\bf r}e^{i{\bf q}\cdot {\bf r}}h({\bf r},t)$ is the Fourier transform of the height fluctuation, whose auto-correlation is 

\begin{equation}
\langle h_{\bf q}(0)h_{{\bf q}^{\prime}}(t)\rangle = 4\pi^2 F(q)\delta({\bf q}+{\bf q}^{\prime})e^{-\omega_qt}  
\label{auto}
\end{equation}

where 
\begin{equation}
F(q)=\frac{k_B T}{\kappa q^4+ \sigma q^2}
\label{fq}
\end{equation}
from equipartition theorem, where $\kappa$ is the bending modulus and $\sigma$ is the surface tension of the membrane.  Assuming an impermeable, flat cell membrane which separates two fluids of mean viscosity $\eta$, the wavelength relaxation rate $\omega_q$ is given by \cite{prost, prost1,siefert}\\

\begin{equation}
    \omega_q=\frac{\kappa q^4 +\sigma q^2}{4\eta q}
\label{omega}    
\end{equation}\\
After using (\ref{auto}) in (\ref{wiener}), and switching to plane polar coordinates, it follows that 

\begin{equation}
PSD_z = \frac{1}{\pi}\int_{q_{\rm min}}^{q_{\rm max}}dq q F(q)\frac{\omega_q}{\omega_q^2+\omega^2}
\label{psd}
\end{equation}

%\begin{equation}
    %PSD=\int_{q_{min}}^{q_{max}}\frac{d^2q}{(2\pi)^2}\int_{-\infty}^{+\infty}\langle h_q(t)h_-q(0)\rangle \exp^{i\omega t} dt
%\end{equation}\\
%Here $h_q$ is the extension of the membrane from the equilibrium position. Simplifying expression(2) gives;\\

%

%\begin{equation}
%    PSD =\int_{q_{min}}^{q_{max}} \langle h^2_q\rangle %\frac{\omega(q)}{\omega(q)^2 + \omega^2}q dq

%\end{equation}

%The parameters $\alpha$ and $\beta$ in expression (4) are the bending modulus and tension on the membrane respectively. Now the expression for Z-power spectral density of a stuck particle on a membrane is,\\

If we consider the cell has an infinite membrane with a point like detection area, $q_{\rm min} =0$ and $ q_{\rm max} =\infty$ in (\ref{psd}). Next, after using (\ref{fq}) and (\ref{omega}) in (\ref{psd}), it follows that the Z-power spectral density of a particle stuck on the membrane is

\begin{equation}
    PSD_z=\frac{4\eta k_{B}T}{\pi}\int_{0}^{\infty}\frac{dq}{(\kappa q^3 +\sigma q)^2 +(4\eta\omega)^2}
\end{equation}\\
In the low frequency limit, ie., when $\omega \rightarrow 0$, it can be shown that
%\begin{equation}
%    PSD_z=\frac{k_{B}T}{4\beta \omega}
%\end{equation}
\\
\begin{equation}
PSD_z\sim \frac{k_{B}T}{2\sigma \omega} ~~~(\omega\to 0), 
\label{psd-small}
\end{equation}
whereas in the large $\omega$ limit, we find 

\begin{equation}
    PSD_z\sim \frac{k_{B}T}{3 (4\eta ^2\kappa)^{1/3}\omega ^{5/3}}~~~(\omega\to\infty).
\label{psd-large}    
\end{equation}

%In the case of high frequency limits that is when  $\omega\rightarrow \infty$\\. We find,
%\begin{equation}
%    PSD_z=\frac{k_{B}T}{12\pi (2\rho ^2\alpha)^{1/3}\omega ^{5/3}}
%\end{equation}

 The expression in (\ref{psd-large}) suggests that the Z-power spectrum obeys a power-law decay at large frequencies $\omega$, with an exponent ${-5/3}$ \\
 
Consider a birefringent particle stuck on the cell membrane, which is characterised by height fluctuations $h({\bf r},t)$, where ${\bf r}=(x,y)$ are points on the plane of projection, which we 
define as the $x-y$ plane.

The slope of the optic axis at a particular instant in the h-r plane is given by,\\
\begin{equation}
    \tan(\theta)=\frac{h_2-h_1}{r_2-r_1}
    \label{theta}
\end{equation}

where the particle touches the cell membrane between $r_1$ and $r_2$, such that $r_2-r_1$ is the length of the contact for the particle. This is of the order of 100 nm for a 1 $\mu$m diameter particle and is assumed to remain constant during rotational motion. 

For small angles $\theta$, we may approximate $\tan~\theta \approx \theta$. Within this approximation, the appropriate generalisation of (\ref{theta}) for the two-dimensional membrane surface is
\begin{equation}
%    \theta=\partial_r(h_q)
\theta({\bf r},t)=\partial_r h({\bf r},t), 
\label{theta1}
\end{equation}
where ${\bf r}$ is the location of the centre of the particle in the $x-y$ plane.  In terms of the Fourier transform $h_{\bf q}(t)$, the angle $\theta$ becomes

\begin{equation}
\theta({\bf r},t)=-\frac{i}{(2\pi)^2}\int d^2{\bf q} h_{\bf q}(t)q\cos\phi e^{-iqr\cos\phi},    
\end{equation}
where $\phi$ is the angle between the (fixed) vector ${\bf r}$ and ${\bf q}$. After using the auto-correlation for the height field given in (\ref{auto}) and carrying out the angular integration, the auto-correlation of the angle becomes

\begin{equation}
\langle \theta({\bf r},0)\theta({\bf r},t)\rangle= \frac{1}{4\pi} \int dq q^3 F(q)e^{-\omega_q t}
\label{auto-theta}
\end{equation}
where the function $F(q)$ has been given in (\ref{fq}). Upon substituting the latter in (\ref{auto-theta}), and using the Wiener-Khinchin theorem, the PSD for the angle/slope fluctuations is found have the general form

\begin{equation}
    PSD_{\theta}=\frac{2\eta k_{B}T}{\pi}\int dq  \frac{q^2}{(\kappa q^3 +\sigma q)^2 +(4\eta\omega)^2}
\label{psd-slope} 
\end{equation}
After a careful analysis of the integral , we find that the low frequency and high frequency behaviours of (\ref{psd-slope}) are given by 

\begin{equation}
 PSD_{\theta}= \frac{8k_B T\eta}{3\pi\sqrt{\sigma^3 \kappa}}-\frac{4k_B T\eta^2}{\sigma^3}\omega~~~~ (\omega\to 0)
\label{result1}
\end{equation}
\begin{equation}
PSD_{\theta}= \frac{k_B T}{12 \kappa\omega}~~~~ (\omega\to \infty) 
\label{result2}
\end{equation}

Thus, we find the functional relationships for the PSD for pitch motion at low and high frequencies. 

%Here $h_q$ is,\\
%\begin{equation}
%    h_q=h_0 e^{-iqr}
%\end{equation}
%The derivative of this expression would give\\
%\begin{equation}
%    \partial_r(h_q)=-iqh_0e^{-iqr}
%\end{equation}\\
%This is the change in slope when the birefringent particle rocke on the cell membrane,\\
%Substituting for $h_q$ with $\theta$ and solving for the slope fluctuations,\\
%\begin{equation}
%    PSD_{slope}=\frac{4\rho kT}{\pi}\int q^2 \frac{1}{(\alpha q^3 +\beta q)^2 +(4\rho\omega)^2}dq
%\end{equation}\\
%At high frequencies,ie when,$  q^3>>q$ and $\omega\rightarrow \infty$\\.
%\begin{equation}
%    PSD_{slope}=\frac{kT}{6\alpha \omega}
%\end{equation}
%Which is the reason why membrane slope power spectrum goes as $\frac{1}{\omega}$.\\

\section{Experimental details}

\begin{figure}[h]
\centering
  \includegraphics[width=\linewidth]{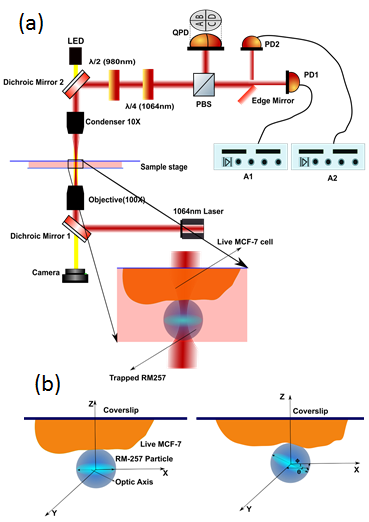}
  \caption{(a) Schematic diagram of the set-up used to detect pitch rotation. A very well polarized 1064 nm laser beam is used to trap the particle, which then passes through into the forward scatter direction. The component of the forward scattered light orthogonal to the input polarization is sent into an edge mirror to ascertain the asmmetry in the scatter pattern. (b) The pitch rotation detection technique is used to find the local slope fluctuations of the cell membrane as shown in this cartoon. }
  \label{schematic}
\end{figure}

The experiment was performed using an optical tweezers kit OTKB/M (Thorlabs, USA) in an inverted configuration, where a linearly polarized 1.7 W, 1064 nm wavelength diode laser (Lasever, China) was used to form the optical tweezers. The objective was an Olympus 100X, 1.3 NA oil immersion one with the illumination aperture being overfilled and the condenser being a 10x, 0.25 NA Nikon air-immersion one. The power of laser light at the sample plane was set to be about 100 mW. The schematic diagram has been shown in Fig. \ref{schematic}. An LED lamp illuminates the sample from the top using a dichroic mirror, while another dichroic collects the visible light to be placed in a CMOS camera (Thorlabs). The forward scattered light emerges through the top dichroic and is sent into a polarizing beam splitter, where most of the light through one of the ports and sent into a Quadrant Photodiode (QPD). The other arm experiences a minimum in scattered intensity and experiences a complete dark when there are no particles in the trapping region. 

The tracer particles that we used are birefringent liquid crystalline RM257 (Merck) particles made using standard techniques \cite{avin,rahul} and of typical diameter 1 $\pm$ 0.1  $\mu$m. When these particles are trapped in optical tweezers, the birefringence axis aligns with that of the polarization of light, both in the conventional yaw and the pitch sense. 
When a well-linearly polarized light is used to trap a birefringent particle, some amount of light also emerges from the dark port of the polarizing beam splitter placed in the forward direction, due to the internal structure of the directors of the particle resulting in a four-lobe scatter intensity pattern. It has been shown in \cite{basudev} that the distribution of light in between these halves becomes anisotropic when the particle turns in the pitch sense. We exploit this very facet to ascertain the pitch motion. 

We place an edge mirror in the path of the dark port of the polarizing beam splitter (PBS) in the forward direction and send one half of the scattered light into one photodiode (PD1) , while sending the other half to a different photodiode (PD2). These photodiode signals are amplified with current amplifiers and then sent into the Data Acquisition System (DAQ card, National Instruments). These time series signals from PD1 and PD2 are then subtracted to gain the pitch signal. The advantage of using this configuration, as opposed to another QPD, is that larger gains can be obtained here. 

\begin{figure}[h]
 \centering
 \includegraphics[width=\linewidth]{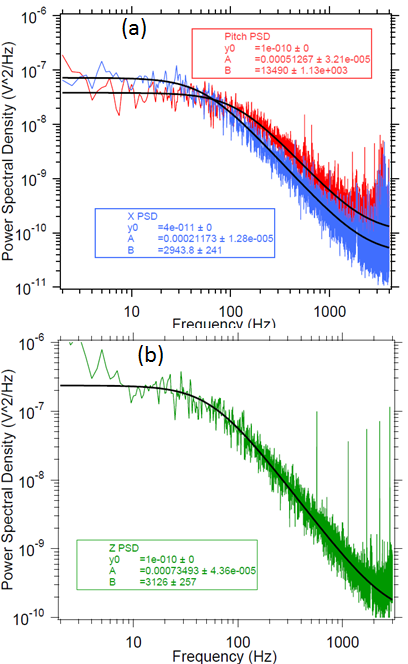}
 \caption{The power spectra for (a) the pitch motion and the transverse x motion (b) for the axial Z motion, fitted to lorentzians (eq. (\ref{lorentz})) for calibration purposes.}
 \label{calibration}
\end{figure}

Typical X, Z and pitch power spectra for a birefringent particle trapped in water are shown in Fig. \ref{calibration}. 

The X and Z PSD are the usual lorentzian in nature, the pitch spectra is also found to be a good Lorentzian. This can be used for calibrating the pitch motion using eq. (\ref{lorentz}) \cite{erik}. 

Michigan Cancer Foundation-7 (MCF-7) cells were grown on glass slides coated with gelatin. These slides were initially treated with the piranha solution and sterilized with a UV (265nm) lamp for 20 minutes and thereafter coated with 0.5\% gelatin solution. MCF7 cells were added towards the center of the coverslip and the Dulbecco's Modified Eagle Medium (DMEM) supplemented with 10\% fetal bovine serum and 1\% glutamine-penicillin-streptomycin was added on top of the coverslip. 10$\mu$L of birefringent sample with particles suspended in water was added to the cells.  Cells were incubated at 5\% carbondioxide and 37 C. 

One such birefringent particle was trapped and gradually brought in contact with the cell surface and held for about 10 seconds. It is observed that the particle attaches to the cell by forming non-specific binding to present us with an excellent opportunity to probe the fluctuations of the cell membrane \cite{rahul1}. We simultaneously probe the slope fluctuations of the membrane from the light scattered by the birefringent particle while in contact with the membrane, as explained in eq. (\ref{result1}). 

\section{Results and discussions}

The power spectral density of the motion of the particle normal to the membrane and the slope fluctuations are reported in Fig. \ref{psd}.  

\begin{figure}[h]
\centering
  \includegraphics[width=\linewidth]{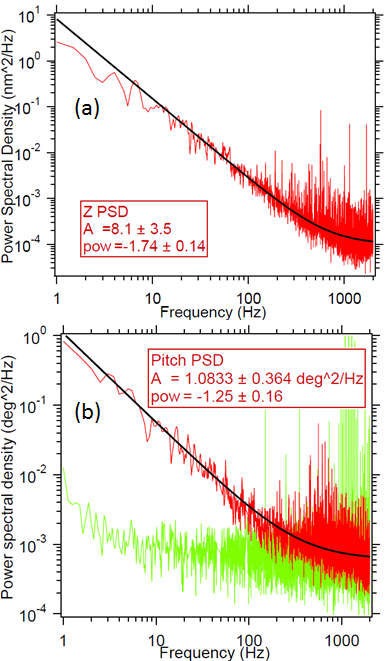}
  \caption{The calibrated Power Spectral Densities (PSD) for (a) the normal fluctuations of the membrane (b) the local slope fluctuations of the membrane indicated by the Pitch angle. In (b), the background PSD with the particle placed on a solid glass surface (without membrane fluctuations) is shown in green. }
  \label{psd}
\end{figure}

The Fig. \ref{psd}(a) indicates the PSD for the normal motion of the cell membrane. We find this to fit well to a power law with exponent -$\frac{5}{3}$  \cite{yu}, particularly at high frequencies between 10 Hz and 1 KHz. This is consistent with the theory presented in eq. (\ref{psd-large}) for normal fluctuations, thereby indicating that the particle is indeed attached to the cell membrane and probing the normal fluctuations. We simultaneously ascertain the slope fluctuation PSD and show in Fig. \ref{psd}(b). This PSD fits well to a power law and shows an exponent of 1.25 $\pm$ 0.16, which we call the pitch PSD. Calibrating the pitch motion amplitude with factors from Fig. \ref{calibration}, we find ourselves capable of resolving 100 mdeg at 40 Hz. We also show the noise floor in Fig. \ref{psd}(b) (green curve). The amplitude of the power law is 1 $\pm$ 0.3 deg$^2$.   

\begin{figure}[htbp]
 \centering
 \includegraphics[width=\linewidth]{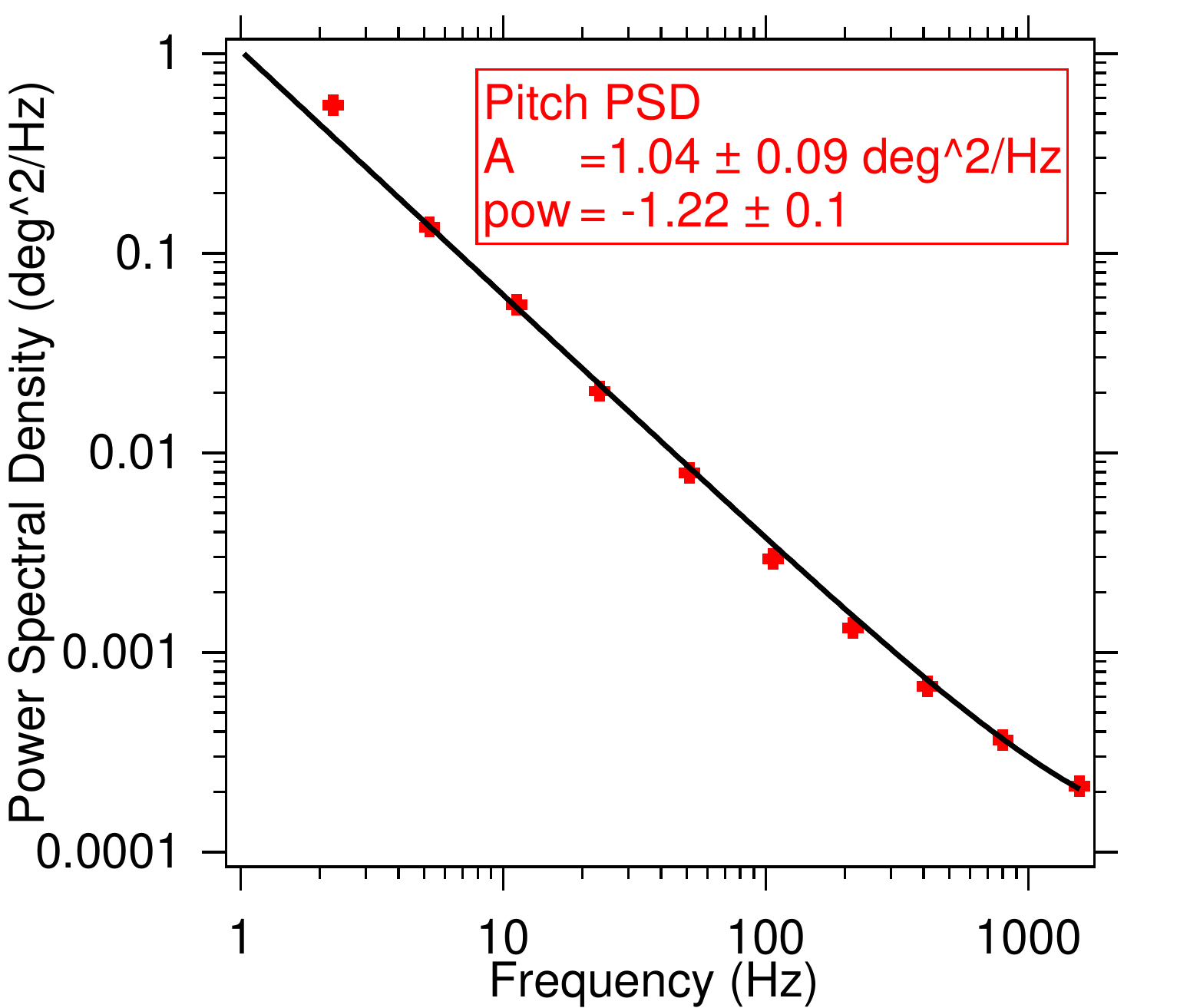}
 \caption{A pitch PSD data for a particle placed on the cell membrane is shown here. The data has been averaged in logarithmic blocks and then subsequently fitted to a power law. The data fits well to 5 \% error above 5 Hz but starts deviating as one goes lower than 5 Hz.}
 \label{block}
\end{figure}

In order to ascertain the accuracy of the power law when fitted to the pitch PSD, we block average \cite{flyvberg} the PSD data in exponents of 2 (namely 1,2,4,8 and so on). This block averaged PSD also fits well to the power law, within 5\% error, till about 5 Hz but starts deviating upon using lower frequencies, with the exponent being 1.22 $\pm$ 0.15, as shown in Fig. \ref{block}.

\begin{figure}[htbp]
 \centering
 \includegraphics[width=\linewidth]{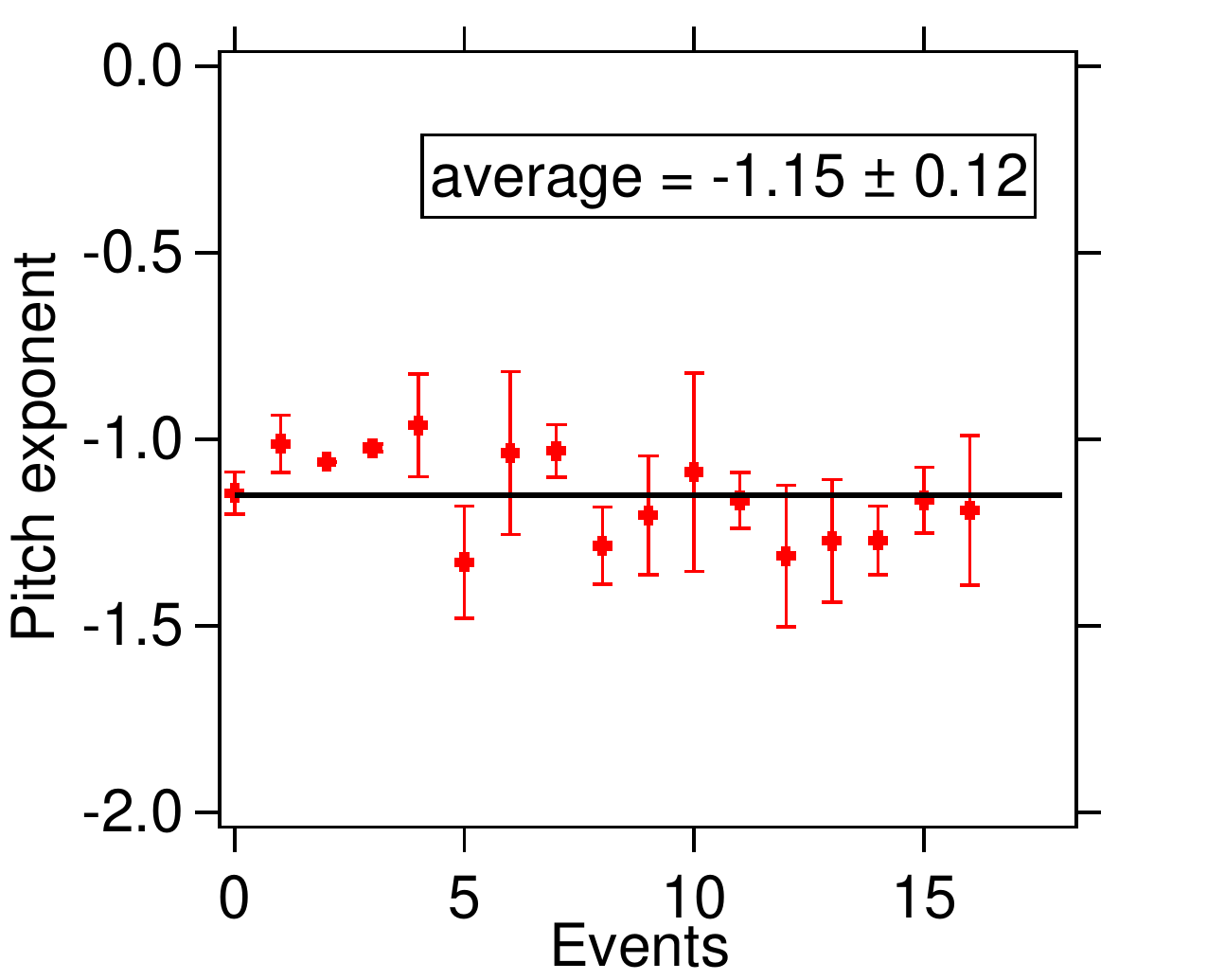}
 \caption{The variation of pitch exponents for different measurement events. The average value of the exponent to the power law fit is -1.15 $\pm$ 0.12, consistent to -1 with a p-value of 0.0001.}
 \label{exponent}
\end{figure}

 We also show the statistics of pitch exponents observed in our experiments in Fig. \ref{exponent}. 

The average value of the pitch exponent is obtained to be -1.15 $\pm$ 0.12. This exponent is comparable with the expected pitch exponent of -1, as indicated in eq. (\ref{result1}), and consistent to a p-value of 0.0001. 

\begin{figure}[htbp]
 \centering
 \includegraphics[width=\linewidth]{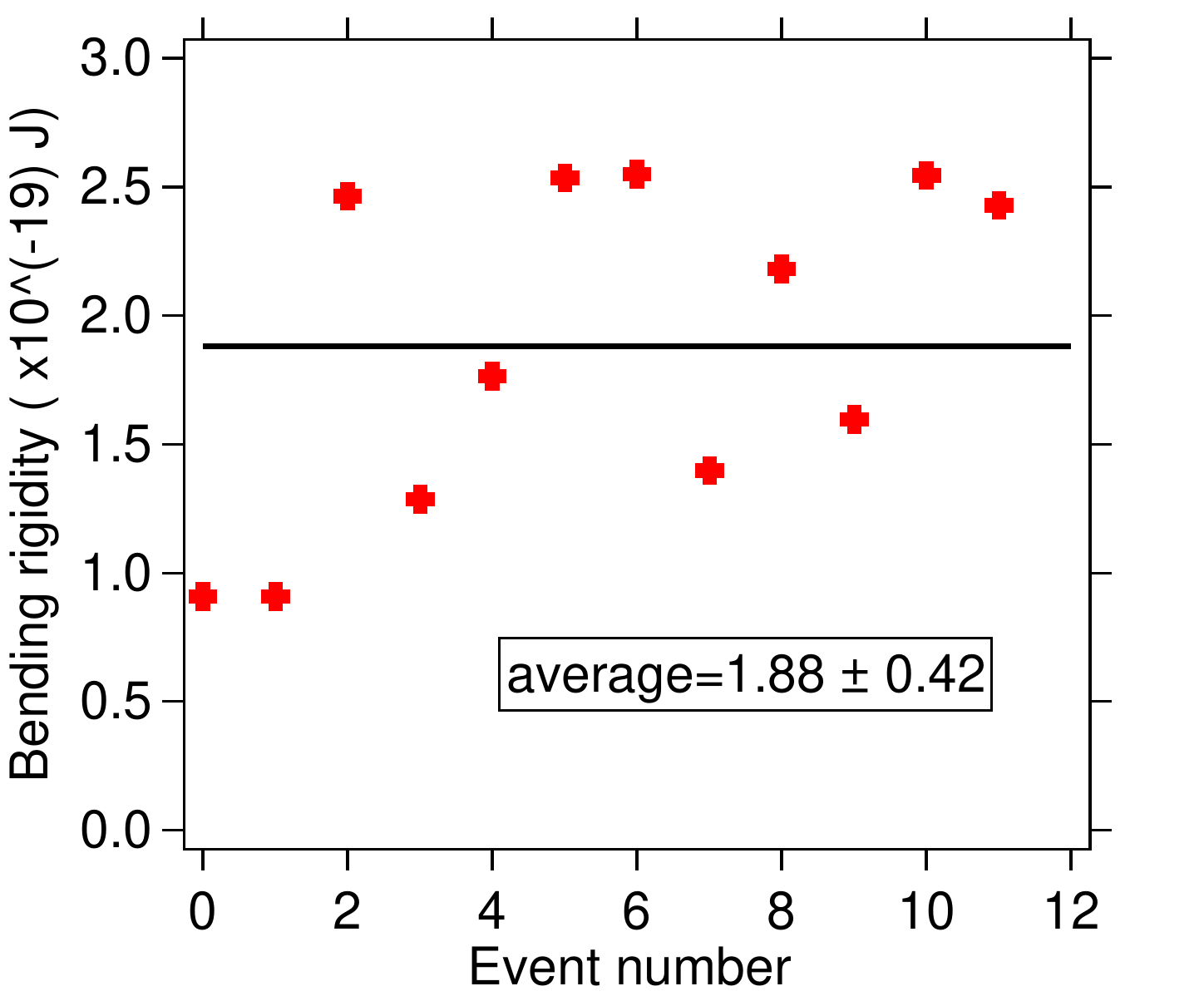}
 \caption{This figure shows the variation of the calculated bending rigidity from the amplitudes of the power laws fitted to the high frequency region of the pitch PSD. The values are consistent with the values previously mentioned in the literature. }
 \label{rigidity}
\end{figure}

We also show, in Fig. \ref{rigidity}, that the value of the bending rigidity estimated from the measurement of the slope is coming to be 1.88 $\pm$ 0.42 $\times$ $10^{-19}$ J, which is consistent with literature values \cite{biswas}. The curves for the normal and slope fluctuations can be simultaneously used to ascertain the bending rigidity, the cytoplasmic viscosity and the surface tension at high accuracy.   

\section{Conclusions}

Thus to conclude, we have developed a new technique to ascertain the pitch rotational motion to a high sensitivity using optical tweezers. The pitch power spectrum for a birefringent particle trapped in water fits well to a Lorentzian. This particle can be attached non-specifically to a cell membrane by holding it against the membrane for 10 seconds. As soon as the particle attaches to the membrane, the vertical fluctuations of the particle can be used to find the membrane fluctuations. The PSD of the vertical fluctuations shows a power law exponent of -$\frac{5}{3}$, confirming that the particle is indeed recording the normal membrane fluctuations. We simultaneously ascertain the slope fluctuations of the membrane and find that the PSD fits well to a power law with the exponent consistent to -1 within 2 standard deviations.  

\section*{Acknowledgements}
We thank the Indian Institute of Technology Madras for their seed and initiation grants. 

%%%END OF MAIN TEXT%%%

%The \balance command can be used to balance the columns on the final page if desired. It should be placed anywhere within the first column of the last page.

\balance

%If notes are included in your references you can change the title from 'References' to 'Notes and references' using the following command:
%\renewcommand\refname{Notes and references}

%%%REFERENCES%%%
\bibliography{rsc} %You need to replace "rsc" on this line with the name of your .bib file

\providecommand*{\mcitethebibliography}{\thebibliography}
\csname @ifundefined\endcsname{endmcitethebibliography}
{\let\endmcitethebibliography\endthebibliography}{}
\begin{mcitethebibliography}{30}
\providecommand*{\natexlab}[1]{#1}
\providecommand*{\mciteSetBstSublistMode}[1]{}
\providecommand*{\mciteSetBstMaxWidthForm}[2]{}
\providecommand*{\mciteBstWouldAddEndPuncttrue}
  {\def\EndOfBibitem{\unskip.}}
\providecommand*{\mciteBstWouldAddEndPunctfalse}
  {\let\EndOfBibitem\relax}
\providecommand*{\mciteSetBstMidEndSepPunct}[3]{}
\providecommand*{\mciteSetBstSublistLabelBeginEnd}[3]{}
\providecommand*{\EndOfBibitem}{}
\mciteSetBstSublistMode{f}
\mciteSetBstMaxWidthForm{subitem}
{(\emph{\alph{mcitesubitemcount}})}
\mciteSetBstSublistLabelBeginEnd{\mcitemaxwidthsubitemform\space}
{\relax}{\relax}

\bibitem[Parsons \emph{et~al.}(2010)Parsons, Horwitz, and Schwartz]{parsons}
J.~T. Parsons, A.~R. Horwitz and M.~A. Schwartz, \emph{Nat. Rev. Mol. Cell
  Biol.}, 2010, \textbf{11}, 633--643\relax
\mciteBstWouldAddEndPuncttrue
\mciteSetBstMidEndSepPunct{\mcitedefaultmidpunct}
{\mcitedefaultendpunct}{\mcitedefaultseppunct}\relax
\EndOfBibitem
\bibitem[Lecuit and Lenne(2007)]{lecuit}
T.~Lecuit and P.~F. Lenne, \emph{Nat. Rev. Mol. Cell Biol.}, 2007, \textbf{8},
  633--644\relax
\mciteBstWouldAddEndPuncttrue
\mciteSetBstMidEndSepPunct{\mcitedefaultmidpunct}
{\mcitedefaultendpunct}{\mcitedefaultseppunct}\relax
\EndOfBibitem
\bibitem[McMahon and Gallop(2005)]{mcmahon}
H.~T. McMahon and J.~T. Gallop, \emph{Nature}, 2005, \textbf{438},
  590--596\relax
\mciteBstWouldAddEndPuncttrue
\mciteSetBstMidEndSepPunct{\mcitedefaultmidpunct}
{\mcitedefaultendpunct}{\mcitedefaultseppunct}\relax
\EndOfBibitem
\bibitem[Kim \emph{et~al.}(2015)Kim, Jo, Hong, Kim, Lee, Heo, and Kim]{kim}
J.~Kim, H.~Jo, H.~Hong, M.~H. Kim, J.~K. Lee, W.~D. Heo and J.~Kim, \emph{Nat.
  Comm.}, 2015, \textbf{6}, 6781\relax
\mciteBstWouldAddEndPuncttrue
\mciteSetBstMidEndSepPunct{\mcitedefaultmidpunct}
{\mcitedefaultendpunct}{\mcitedefaultseppunct}\relax
\EndOfBibitem
\bibitem[Park \emph{et~al.}(2008)Park, Diez-Silva, Popescu, Lykotrafitis, Choi,
  Feld, and Suresh]{park}
Y.~Park, M.~Diez-Silva, G.~Popescu, G.~Lykotrafitis, W.~Choi, M.~S. Feld and
  S.~Suresh, \emph{Proc. Nat. Acad. Sci. (USA)}, 2008, \textbf{105},
  13730--13735\relax
\mciteBstWouldAddEndPuncttrue
\mciteSetBstMidEndSepPunct{\mcitedefaultmidpunct}
{\mcitedefaultendpunct}{\mcitedefaultseppunct}\relax
\EndOfBibitem
\bibitem[Connes \emph{et~al.}(2016)Connes, Alexy, Detterich, Romana,
  Hardy-Dessources, and Ballas]{connes}
P.~Connes, T.~Alexy, J.~Detterich, M.~Romana, M.-D. Hardy-Dessources and S.~K.
  Ballas, \emph{Blood Rev.}, 2016, \textbf{30}, 111--118\relax
\mciteBstWouldAddEndPuncttrue
\mciteSetBstMidEndSepPunct{\mcitedefaultmidpunct}
{\mcitedefaultendpunct}{\mcitedefaultseppunct}\relax
\EndOfBibitem
\bibitem[Wirtz \emph{et~al.}(2011)Wirtz, Konstantopoulos, and Searson]{wirtz}
D.~Wirtz, K.~Konstantopoulos and P.~C. Searson, \emph{Nat. Rev. Cancer}, 2011,
  \textbf{11}, 512--522\relax
\mciteBstWouldAddEndPuncttrue
\mciteSetBstMidEndSepPunct{\mcitedefaultmidpunct}
{\mcitedefaultendpunct}{\mcitedefaultseppunct}\relax
\EndOfBibitem
\bibitem[Winograd-Katz \emph{et~al.}(2014)Winograd-Katz, Fassler, Geiger, and
  Legate]{katz}
S.~E. Winograd-Katz, R.~Fassler, B.~Geiger and K.~R. Legate, \emph{Nat. Rev.
  Mol. Cell Biol.}, 2014, \textbf{15}, 273--288\relax
\mciteBstWouldAddEndPuncttrue
\mciteSetBstMidEndSepPunct{\mcitedefaultmidpunct}
{\mcitedefaultendpunct}{\mcitedefaultseppunct}\relax
\EndOfBibitem
\bibitem[Gov \emph{et~al.}(2003)Gov, Zilman, and Safran]{gov}
N.~Gov, A.~G. Zilman and S.~Safran, \emph{Phys. Rev. Lett.}, 2003, \textbf{90},
  228101\relax
\mciteBstWouldAddEndPuncttrue
\mciteSetBstMidEndSepPunct{\mcitedefaultmidpunct}
{\mcitedefaultendpunct}{\mcitedefaultseppunct}\relax
\EndOfBibitem
\bibitem[Gov and Safran(2005)]{gov1}
N.~S. Gov and S.~A. Safran, \emph{Biophys. J.}, 2005, \textbf{88},
  1859--1874\relax
\mciteBstWouldAddEndPuncttrue
\mciteSetBstMidEndSepPunct{\mcitedefaultmidpunct}
{\mcitedefaultendpunct}{\mcitedefaultseppunct}\relax
\EndOfBibitem
\bibitem[Park \emph{et~al.}(2010)Park, Best, Badizadegan, Dasari, Feld,
  Kuriabova, Henle, Levine, and Popescu]{park1}
Y.~Park, C.~A. Best, K.~Badizadegan, R.~R. Dasari, M.~S. Feld, T.~Kuriabova,
  M.~L. Henle, A.~J. Levine and G.~Popescu, \emph{Proc. Nat. Acad. Sci. (USA)},
  2010, \textbf{107}, 6731--6736\relax
\mciteBstWouldAddEndPuncttrue
\mciteSetBstMidEndSepPunct{\mcitedefaultmidpunct}
{\mcitedefaultendpunct}{\mcitedefaultseppunct}\relax
\EndOfBibitem
\bibitem[Yu \emph{et~al.}(2018)Yu, Yang, Yang, Zhang, Wang, and Tao]{yu}
H.~Yu, Y.~Yang, Y.~Yang, F.~Zhang, S.~Wang and N.~Tao, \emph{Nanoscale}, 2018,
  \textbf{10}, 5133--5139\relax
\mciteBstWouldAddEndPuncttrue
\mciteSetBstMidEndSepPunct{\mcitedefaultmidpunct}
{\mcitedefaultendpunct}{\mcitedefaultseppunct}\relax
\EndOfBibitem
\bibitem[Biswas \emph{et~al.}(2017)Biswas, Alex, and Sinha]{biswas}
A.~Biswas, A.~Alex and B.~Sinha, \emph{Biophs. J.}, 2017, \textbf{113},
  1768--1781\relax
\mciteBstWouldAddEndPuncttrue
\mciteSetBstMidEndSepPunct{\mcitedefaultmidpunct}
{\mcitedefaultendpunct}{\mcitedefaultseppunct}\relax
\EndOfBibitem
\bibitem[Giannone \emph{et~al.}(2004)Giannone, Dubin-Thaler, Dobereiner,
  Kieffer, Bresnick, and Sheetz]{giannone}
G.~Giannone, B.~J. Dubin-Thaler, H.-G. Dobereiner, N.~Kieffer, A.~R. Bresnick
  and M.~P. Sheetz, \emph{Cell}, 2004, \textbf{116}, 441--443\relax
\mciteBstWouldAddEndPuncttrue
\mciteSetBstMidEndSepPunct{\mcitedefaultmidpunct}
{\mcitedefaultendpunct}{\mcitedefaultseppunct}\relax
\EndOfBibitem
\bibitem[Dobereiner \emph{et~al.}(2006)Dobereiner, Dubin-Thaler, Hofman,
  Xenias, Sims, Giannone, Dustin, Wiggins, and Sheetz]{dobereiner}
H.-G. Dobereiner, B.~J. Dubin-Thaler, J.~M. Hofman, H.~S. Xenias, T.~N. Sims,
  G.~Giannone, M.~L. Dustin, C.~H. Wiggins and M.~P. Sheetz, \emph{Phys. Rev.
  Lett.}, 2006, \textbf{97}, 038102\relax
\mciteBstWouldAddEndPuncttrue
\mciteSetBstMidEndSepPunct{\mcitedefaultmidpunct}
{\mcitedefaultendpunct}{\mcitedefaultseppunct}\relax
\EndOfBibitem
\bibitem[Chen \emph{et~al.}(2009)Chen, Tsai, Wang, and Lee]{chen}
C.-H. Chen, F.-C. Tsai, C.-C. Wang and C.-H. Lee, \emph{Phys. Rev. Lett.},
  2009, \textbf{103}, 238101\relax
\mciteBstWouldAddEndPuncttrue
\mciteSetBstMidEndSepPunct{\mcitedefaultmidpunct}
{\mcitedefaultendpunct}{\mcitedefaultseppunct}\relax
\EndOfBibitem
\bibitem[Monzel \emph{et~al.}(2015)Monzel, Schmidt, Kleusch, Kirchenbuchler,
  Seifert, Smith, Sengupta, and Merkel]{monzel}
C.~Monzel, D.~Schmidt, C.~Kleusch, D.~Kirchenbuchler, U.~Seifert, A.-S. Smith,
  K.~Sengupta and R.~Merkel, \emph{Nat. Commun.}, 2015, \textbf{6}, 8162\relax
\mciteBstWouldAddEndPuncttrue
\mciteSetBstMidEndSepPunct{\mcitedefaultmidpunct}
{\mcitedefaultendpunct}{\mcitedefaultseppunct}\relax
\EndOfBibitem
\bibitem[Santos \emph{et~al.}(2016)Santos, Deturche, Vezy, and Jaffiol]{santos}
M.~C.~D. Santos, R.~Deturche, C.~Vezy and R.~Jaffiol, \emph{Biophys. J.}, 2016,
  \textbf{111}, 1316--1327\relax
\mciteBstWouldAddEndPuncttrue
\mciteSetBstMidEndSepPunct{\mcitedefaultmidpunct}
{\mcitedefaultendpunct}{\mcitedefaultseppunct}\relax
\EndOfBibitem
\bibitem[Florin \emph{et~al.}(1997)Florin, Pralle, Horber, and Stelzer]{florin}
E.-L. Florin, A.~Pralle, J.~K.~H. Horber and E.~H.~K. Stelzer, \emph{J. Struct.
  Biol.}, 1997, \textbf{119}, 202--211\relax
\mciteBstWouldAddEndPuncttrue
\mciteSetBstMidEndSepPunct{\mcitedefaultmidpunct}
{\mcitedefaultendpunct}{\mcitedefaultseppunct}\relax
\EndOfBibitem
\bibitem[Friedrich and Rohrbach(2015)]{rohrbach}
L.~Friedrich and A.~Rohrbach, \emph{Nat. Nano.}, 2015, \textbf{10},
  1064--1069\relax
\mciteBstWouldAddEndPuncttrue
\mciteSetBstMidEndSepPunct{\mcitedefaultmidpunct}
{\mcitedefaultendpunct}{\mcitedefaultseppunct}\relax
\EndOfBibitem
\bibitem[Junger \emph{et~al.}(2015)Junger, Kohler, Meinel, Meyer, Nitschke,
  Erhard, and Rohrbach]{rohrbach1}
F.~Junger, F.~Kohler, A.~Meinel, T.~Meyer, R.~Nitschke, B.~Erhard and
  A.~Rohrbach, \emph{Biophys. J.}, 2015, \textbf{109}, 869--882\relax
\mciteBstWouldAddEndPuncttrue
\mciteSetBstMidEndSepPunct{\mcitedefaultmidpunct}
{\mcitedefaultendpunct}{\mcitedefaultseppunct}\relax
\EndOfBibitem
\bibitem[Roy \emph{et~al.}(2018)Roy, Ramaiya, and Schaffer]{basudev}
B.~Roy, A.~Ramaiya and E.~Schaffer, \emph{J. Opt.}, 2018, \textbf{20},
  035603\relax
\mciteBstWouldAddEndPuncttrue
\mciteSetBstMidEndSepPunct{\mcitedefaultmidpunct}
{\mcitedefaultendpunct}{\mcitedefaultseppunct}\relax
\EndOfBibitem
\bibitem[Schaffer \emph{et~al.}(2007)Schaffer, Norrelykke, and Howard]{erik}
E.~Schaffer, S.~F. Norrelykke and J.~Howard, \emph{Langmuir}, 2007,
  \textbf{23}, 3654--3665\relax
\mciteBstWouldAddEndPuncttrue
\mciteSetBstMidEndSepPunct{\mcitedefaultmidpunct}
{\mcitedefaultendpunct}{\mcitedefaultseppunct}\relax
\EndOfBibitem
\bibitem[Ramaswamy \emph{et~al.}(1999)Ramaswamy, Toner, and Prost]{prost}
S.~Ramaswamy, J.~Toner and J.~Prost, \emph{Pramana}, 1999, \textbf{53},
  237--242\relax
\mciteBstWouldAddEndPuncttrue
\mciteSetBstMidEndSepPunct{\mcitedefaultmidpunct}
{\mcitedefaultendpunct}{\mcitedefaultseppunct}\relax
\EndOfBibitem
\bibitem[Prost \emph{et~al.}(1998)Prost, Manneville, and Bruinsma]{prost1}
J.~Prost, J.-B. Manneville and R.~Bruinsma, \emph{Eur. Phys. J. B}, 1998,
  \textbf{1}, 465--480\relax
\mciteBstWouldAddEndPuncttrue
\mciteSetBstMidEndSepPunct{\mcitedefaultmidpunct}
{\mcitedefaultendpunct}{\mcitedefaultseppunct}\relax
\EndOfBibitem
\bibitem[Siefert(1994)]{siefert}
U.~Siefert, \emph{Phys. Rev. E}, 1994, \textbf{49}, 3124--3127\relax
\mciteBstWouldAddEndPuncttrue
\mciteSetBstMidEndSepPunct{\mcitedefaultmidpunct}
{\mcitedefaultendpunct}{\mcitedefaultseppunct}\relax
\EndOfBibitem
\bibitem[Ramaiya \emph{et~al.}(2017)Ramaiya, Roy, Bugiel, and Schaffer]{avin}
A.~Ramaiya, B.~Roy, M.~Bugiel and E.~Schaffer, \emph{Proc. Nat. Acad. Sci.
  (USA)}, 2017, \textbf{114}, 10894--10899\relax
\mciteBstWouldAddEndPuncttrue
\mciteSetBstMidEndSepPunct{\mcitedefaultmidpunct}
{\mcitedefaultendpunct}{\mcitedefaultseppunct}\relax
\EndOfBibitem
\bibitem[Vaippully \emph{et~al.}(2019)Vaippully, Bhatt, Ranjan, and Roy]{rahul}
R.~Vaippully, D.~Bhatt, A.~D. Ranjan and B.~Roy, \emph{Phys. Scr.}, 2019,
  \textbf{94}, 105008\relax
\mciteBstWouldAddEndPuncttrue
\mciteSetBstMidEndSepPunct{\mcitedefaultmidpunct}
{\mcitedefaultendpunct}{\mcitedefaultseppunct}\relax
\EndOfBibitem
\bibitem[Vaippully \emph{et~al.}(2020)Vaippully, Ramanujan, Bajpai, and
  Roy]{rahul1}
R.~Vaippully, V.~Ramanujan, S.~Bajpai and B.~Roy, \emph{J. Phys. Cond. Mat.},
  2020, \textbf{32}, 235101\relax
\mciteBstWouldAddEndPuncttrue
\mciteSetBstMidEndSepPunct{\mcitedefaultmidpunct}
{\mcitedefaultendpunct}{\mcitedefaultseppunct}\relax
\EndOfBibitem
\bibitem[Berg-Sorensen and Flyvberg(2004)]{flyvberg}
K.~Berg-Sorensen and H.~Flyvberg, \emph{Rev. Sci. Instrum.}, 2004, \textbf{75},
  594--612\relax
\mciteBstWouldAddEndPuncttrue
\mciteSetBstMidEndSepPunct{\mcitedefaultmidpunct}
{\mcitedefaultendpunct}{\mcitedefaultseppunct}\relax
\EndOfBibitem
\end{mcitethebibliography}
\bibliographystyle{rsc} %the RSC's .bst file

\end{document}